# Colossal magnetoresistance in Ti lightly-doped $Cr_2Se_3$ single crystals with layered structure

Shu-Juan Zhang,[†,‡] Jian-Min Yan,[§] F. Tang,[#] Jin Wu,[†] Wei-Qi Dong,[†] Dan-Wen Zhang,[†] Fu-Sheng Luo,[†] Lei Chen,[†] Y. Fang,[#] Tao Zhang,[⊥] Yang Chai,[§] Weiyao Zhao,[*,&] Xiaolin Wang,[&] and Ren-Kui Zheng[*,†]

[†] *School of Materials Science and Engineering, Jiangxi Engineering Laboratory for Advanced Functional Thin Films, Nanchang University, Nanchang 330031, China*

[‡] *School of Materials, Mechanic, and Electrical Engineering, Jiangxi Science and Technology Normal University, Nanchang 330038, China*

[§] *Department of Applied Physics, The Hong Kong Polytechnic University, Hong Kong, China*

[#] *Jiangsu Laboratory of Advanced Functional Materials, Department of Physics, Changshu Institute of Technology, Changshu 215500, China*

[⊥] *School of Physics and Materials Science, Guangzhou University, Guangzhou 510006, China*

[&] *Institute for Superconducting and Electronic Materials, & ARC Centre of Excellence in Future Low-Energy Electronics Technologies, Innovation Campus, University of Wollongong, North Wollongong NSW 2500, Australia*

**ABSTRACT:** Stoichiometric $Cr_2Se_3$ single crystals are particular layer-structured antiferromagnets which possess noncolinear spin configuration, weak ferromagnetic moments, moderate magnetoresistance (MR~14.3%), and bad metallic conductivity below the antiferromagnetic phase transition temperature. Here, we report an interesting >16000% colossal magnetoresistance (CMR) effect in Ti (1.5 atomic percent) lightly-doped $Cr_2Se_3$ single crystals. Such a CMR is approximately 1143 times larger than that of the stoichiometric $Cr_2Se_3$ crystals and is rarely observed in layered antiferromagnets and is attributed to the frustrated spin configuration. Moreover, the Ti doping not only dramatically changes the electronic conductivity of the $Cr_2Se_3$ crystal from a bad metal to a semiconductor with a gap of ~ 15 meV, but also induces a change of the magnetic anisotropy of the $Cr_2Se_3$ crystal from strong out-of-plane to weak in plane. Further, magnetotransport measurements reveal that the low-field MR scales with the square of the reduced magnetization, which is a signature of CMR materials. The layered Ti:Cr$_2$Se$_3$ with CMR effect could be used as 2D heterostructure building blocks to provide colossal negative MR in spintronic devices.

**1. INTRODUCTION**

The magnetism of materials is carried by electron spin which offers a new degree of freedom of electronic transport properties, therefore providing more applications in spintronic devices. One of the most important electronic transport behaviors in magnetic materials (e.g., magnetic multilayers or cluster-alloys) is the giant magnetoresistance (GMR) [1-3] which shows large negative magnetoresistance (MR) under applied magnetic fields. GMR has proved its advantage in applications in hard disk drives, biosensors, microelectromechanical systems and other devices [4]. Around the new millennium, it became recognized that transition metal oxide materials of $R_{1-x}A_x$MnO$_3$ (where $R$ = rare-earth elements, e.g., La, Nd; $A$=alkaline-earth elements, e.g., Ca, Sr, Ba) show larger negative MR, namely, the colossal magnetoresistance (CMR). The CMR effect are usually found near the metal-to-insulator transition during which the disorder of local spins together with local lattice distortion (i.e., magnetic polarons) serves as one of the sources of carrier's localization.

MR is usually expressed as a function of the magnetic field $B$ at a certain temperature $T$: $\text{MR}(B) = [R(B) - R(0)]/R(0)$, where $R(B)$ and $R(0)$ are the resistance of a material in the presence and absence of the magnetic field $B$, respectively. In most cases, MR values of materials are between moderate negative MR to giant or extremely large positive MR (-50% - $1\times10^7$%), where MR and resistance change $\Delta R = R(B) - R(0)$ show similarly shaped curves with $B$. However, in the giant negative MR (CMR included) realm, where $R(B) \ll R(0)$, MR is no longer a good representative of the negative resistance correction phenomenon. In this case, $\text{CMR}(B) = [R(0) - R(B)]/R(B)$ is introduced to highlight the dramatical resistance decrease with applied magnetic fields. One of the famous CMR material family is $R_{1-x}A_x\text{MnO}_3$ whose maximum CMR of a few hundred percent could be achieved around the metal-to-insulator phase transition region [5-7]. CMR (or giant negative MR) family has been expanded into more compounds: antiferromagnetic oxide $\text{NaCr}_2\text{O}_4$ (CMR ~ 5000%) [8], diluted magnetic semiconductors Mn-doped ZnO (CMR ~ 225%) [9] Cr-doped $\text{Ti}_2\text{O}_3$ (CMR ~ 365%) [10] and Mn-doped GaAs (CMR ~$3.2\times10^4$%) [11], rare-earth based compounds $\text{Eu}_{0.99}\text{Gd}_{0.01}\text{Se}$ (CMR ~ $5\times10^6$%) [12], $\text{GdI}_2$ (CMR ~ 230%) [13], EuTe (CMR ~ $10^5$%) [14], $\text{Eu}_{14}\text{MnBi}_{11}$ (CMR ~200 %) [15], and $\text{Eu}_5\text{In}_2\text{Sb}_6$ (CMR ~$1\times10^7$) [16]. According to the existing systems, one may conclude that the materials exhibiting CMR effect should sit on the edge of magnetic and nonmagnetic and/or the edge of insulator and metal.

Based on the principle of searching for CMR materials, we focus on a particular rhombohedral layered material system: chromium selenide $\text{Cr}_2\text{Se}_3$. $\text{Cr}_2\text{Se}_3$ crystallizes in a Cr-vacant NiAs-type lattice where the vacant chromium sites are ordered, leading to very delicate electronic and magnetic structure depending on the Cr concentration. As a candidate thermoelectric material, polycrystal $\text{Cr}_2\text{Se}_3$ shows a figure of merit $ZT$ ~0.25 at ~600 K [17-18]. The neutron powder diffraction analyses showed that the $\text{Cr}_2\text{Se}_3$ possesses complicated antiferromagnetic magnetic structures below the Néel temperature. Specifically, there is a collinear antiferromagnetic state just below $T_\text{N}$ and a noncollinear antiferromagnetic state together with a monoclinic magnetic unit cell at lower temperatures [19]. Guo *et al* [17]. reported that both the paramagnetic fitted effective magnetic moments and the temperature dependent magnetization behaviors change significantly with a very small variation in $x$ for $\text{Cr}_{2+x}\text{Se}_3$ polycrystalline samples. The slightly non-stoichiometric $\text{Cr}_{2.04}\text{Se}_3$ single crystals has been reported as a two-dimensional magnetic semiconductor with a band gap ~8 meV and a significant CMR ~ 300% at $T$=2 K and anomalous Hall effect at low temperatures [20]. Our recent investigation of perfect stoichiometric $Cr_2Se_3$ single crystals show that it has a bad metallic ground state with canted antiferromagnetic magnetic structure below $T_N$ ~ 60 K, below which a moderate negative CMR ~ 14.3% appears under $B$=9 T and 2 K [21]. Due to the tunability of magnetic, electronic, and MR properties by Cr concentration, we deduce that the $Cr_2Se_3$ could be a good system to explore the coupling between conduction carriers and the spin configuration.

In this paper, we employ the chemical vapor transport (CVT) method to grow high quality 1.5 at% Ti-doped $Cr_2Se_3$ (Ti:$Cr_2Se_3$) (1.5 at% is the nomination doping ratio) single crystals and investigated the electronic transport and magnetic properties of the crystals. Magnetotransport and magnetic measurements show that a small amount of Ti doping not only induces an unprecedented negative CMR~$10^4$% in the Ti:$Cr_2Se_3$, but also dramatically changes the electronic conductivity behaviors and magnetic anisotropy, as discussed in the sections that follow. Our results demonstrate that the Ti-doping is a highly effective approach to induce negative CMR effect in the layered $Cr_2Se_3$.

## 2. EXPERIMENTAL SECTION

High-quality Ti:$Cr_2Se_3$ single crystals were grown by the CVT method using the iodine ($I_2$) as a transport agent. Briefly, high-purity Ti, Cr, and Se powders (200 mesh) with a nomination molar ratio of 0.03:1.97:3 and a total mass of ~1 g, together with 20 mg/ml iodine, were sealed in a vacuum ampoule as the starting materials. The crystal growth was carried out in a horizontal two-zone furnace between 950 °C (source) and 850 °C (sink) for one week and furnace cooled to room temperature. Plane-shaped single crystals with a typical size of ~3×5×0.5 mm$^3$ were obtained, as shown in the inset of Figure 1a.

The crystallographic properties of the Ti:Cr$_2$Se$_3$ crystals were characterized using a Rigaku SmartLab x-ray diffractometer equipped with Cu$_{k\alpha 1}$ radiation and a Tecnai G2 F20 S-Twin transmission electron microscope. The chemical composition and element mapping were measured using an energy dispersive x-ray spectrometer (EDS) (Oxford AztecLive UltimMax 80) installed on a FEI focused ion beam system (Thermo Fisher Scientific, Scios2). Raman spectra were recorded on a Raman spectrometer equipped with a 532 nm laser (Witec Alpha300R) as an excitation source at room temperature.

Ohmic contacts were prepared on fresh surface of Ti:Cr$_2$Se$_3$ single crystals using room-temperature cured silver paste. The magnetotransport and magnetic properties were measured in two different configurations: 1) the direction of the magnetic field perpendicular to the *ab* plane of the crystals, and 2) the direction of the magnetic field parallel to the *ab* plane of the crystals, using a physical property measurement system (DynaCool-14, Quantum Design) and a magnetic property measurement system (MPMS3, Quantum Design), respectively.

## 3. RESULTS AND DISCUSSION

The CVT crystal growth process employed here is slightly different from that employed for growing the metallic stoichiometric Cr$_2$Se$_3$ single crystals which were grown from a mixture of Cr and Se powders with 5:4 molar ratio [21]. Here, starting powders with the stoichiometric ratio of 0.03:1.97:3 was employed to grow Ti$_{0.03}$Cr$_{1.97}$Se$_3$ single crystals. After vapor-phase transport, the atomic ratio of obtained single crystals determined by EDS is Ti$_{0.023}$Cr$_{2.002}$Se$_{3.000}$ (Figure 1d). Figures 1e-g display the Cr, Se, and Ti element mapping taken on the *ab* plane of the crystal. The results reveal uniform distribution of the Cr, Se, and Ti elements. Room-temperature Raman scattering measurements were performed on the *ab* plane of a fresh single crystal. The results are shown in Figure 1a. The Cr$_{m1}$ – Cr$_{m5}$ vibrational modes at 304, 342, 536, 596, 700 cm$^{-1}$ agree well with previous reports [21-22]. The x-ray diffraction (XRD) $\theta$-$2\theta$ scan patterns taken on the fresh surface of the crystal is shown in Figure 1b in which the *y*-axis is plotted in a logarithmic scale in order to show the relatively weak (003) and (009) diffraction peaks. There's no obvious peak shifting in the XRD pattern, as compared with that of the stoichiometric $Cr_2Se_3$ single crystals [21], which indicates neglectable lattice distortion upon a very small amount of Ti doping (1.5 at%). The XRD rocking curve measurements taken on the (00<u>12</u>) diffraction peak yields a very narrow full width at half maximum (FWHM) of ~0.04°, as shown in Figure 1c. A high-resolution transmission electron microscopy (HRTEM) image taken on the *M*-plane of a Ti:$Cr_2Se_3$ crystal is shown in Figure 1h from which clear atomic sites and a lattice spacing of $d_{003}$=5.8 Å and the *c*-axis lattice constant of ~17.4 Å can be obtained, consistent with the layered structure of the $Cr_2Se_3$ [19] (Inset of Figure 1a). The sharp selection-area-electron-diffraction (SAED) pattern taken on the *M*-plane further verifies the high crystallinity of the crystal (Figure 1i). All these results demonstrate that the Ti:$Cr_2Se_3$ crystal is single phase with the *c* axis normal to the crystal's surface and has an excellent crystalline quality, which allows us to explore its intrinsic electronic transport and magnetic properties.

For the $Cr_{2-\delta}Se_3$ compounds, the Cr concentration is a key factor to determine the magnetic and electronic transport properties. The stoichiometric $Cr_2Se_3$ crystals show bad metallic conductivity behaviors ($R_{300\ K}/R_{3\ K}$ ~ 2) [21] while the $Cr_{0.68}$Se (corresponding to $Cr_{2.04}Se_3$) crystals show insulating conductivity behaviors with a band gap ~8 meV [20]. Since the composition of the Ti:$Cr_2Se_3$ crystal is determined to be $Ti_{0.023}Cr_{2.002}Se_{3.000}$ which is slightly Cr- and Ti-rich, as compared with the stoichiometric $Cr_2Se_3$, one should expect similar insulating conductivity behaviors for such a crystal. We thus conducted the magnetic and electronic transport measurements on the Ti:$Cr_2Se_3$ single crystal and show the results in Figure 2. The stoichiometric $Cr_2Se_3$ crystal is a noncolinear antiferromagnet below $T_N$ ~ 60 K, exhibiting weak ferromagnetism (WFM). For the Ti:Cr$_2$Se$_3$ crystals, the WFM components also dominate the magnetic behaviors, as shown in Figure 2a in which the magnetization increases with further cooling from the Néel temperature, instead of showing a peak at $T_N$. Let's focus on the in-plane magnetization versus temperature (*M-T*) curves with an applied magnetic field of 0.05 T. There is another magnetic transition (probably from the high-temperature antiferromagnetic phase to low-temperature antiferromagnetic phase [19]) at ~ 42 K. This transition is only observed in the *ab* plane and can be slightly shifted to lower temperature by increasing applied magnetic fields. Another feature of the *M-T* curves is the splitting of the ZFC and FC curves at low temperatures, which is similar to those of the Cr$_2$Se$_3$ [19] and rare earth orthoferrites with WFM [23,24].

Although there are similarities between the magnetic properties of the Ti:Cr$_2$Se$_3$ crystals and the stoichiometric Cr$_2$Se$_3$ crystals [21], there are some differences which should be mentioned: 1) the magnetization of the Ti:Cr$_2$Se$_3$ crystal is one order smaller than that of the Cr$_2$Se$_3$ crystal under the same condition, e.g., the FC magnetization at *T*=10 K and *B*=0.2 T is only 0.035 $\mu_B$/f.u. (0.57 emu/g) for the Ti:Cr$_2$Se$_3$ crystal while 0.5 $\mu_B$/f.u. (8.1 emu/g) for the stoichiometric Cr$_2$Se$_3$ crystal; 2) the magnetic easy axis is along the *c*-axis direction for the stoichiometric Cr$_2$Se$_3$ crystals whose out-of-plane magnetization is approximately 15 times larger than the in-plane one, however, for the Ti-doped Cr$_2$Se$_3$ crystals, the out-of-plane magnetization is only 71% of the in-plane one. Namely, upon Ti doping, the magnetic anisotropy changes from strong out-of-plane direction to weak in-plane direction.

Further, we conducted temperature dependent resistance measurements on the Ti:Cr$_2$Se$_3$ crystal under magnetic fields with their direction parallel to the *ab* plane and perpendicular to the electric current direction and show the results in log-log scale in Figure 2c. One may notice that the zero-field resistance decreases monotonically with heating from 2 to 300 K, which is a feature of band gap opening. In the inset of Figure 2c, we plot the ln$R$ vs. inverse temperature to analyze this behavior following the Arrhenius equation $lnR = lnR_0 + E_g/2k_BT$. A fitting of the data yields the band gap $E_g$=15 meV. For the 2 at% Cr-rich $Cr_{2.04}Se_3$ crystals [20], a band gap of ~ 8 meV was obtained. It is reasonable to observe the semiconducting electronic transport behaviors of the Ti:$Cr_2Se_3$ crystals with a band gap $E_g$=15 meV. Upon the application of magnetic fields, the resistance of the Ti:$Cr_2Se_3$ crystal decreases dramatically below $T_N$, showing CMR effect. We therefore plot the CMR value as a function of temperature under various magnetic fields in Figure 2d. The CMR at a given magnetic field decreases monotonically upon heating and is neglectable (< 10%) above $T_N$ and below 7 T. The sizeable CMR are obtained at low temperatures (< 30 K), where the WFM components dominates the magnetic behaviors. In Figure 2e, we show the magnetic field dependent CMR effect in the 2 – 300 K temperature region in which the field dependent of CMR is much stronger at lower temperatures. Another point in Figure 2e one should notice is that the CMR increase nearly linearly with the magnetic field in high field region (e.g., >2.5 T), which is interesting and desires theoretical models.

To study the magnetotransport anisotropy properties of the Ti:$Cr_2Se_3$ crystal, we further conducted magnetotransport measurements on the crystal with the directions of the magnetic field perpendicular to the *ab* plane and along the electric current direction (Figure S1b,c, Supporting Information), respectively. The corresponding MR versus temperature curves are shown in Figure S1e,f (Supporting Information). The magnetotransport behaviors and the maximum MR values at 2 K (Figure S1a,b,c, Supporting Information) are almost similar for all measurement configurations, demonstrating weak magnetotransport anisotropy of the Ti:$Cr_2Se_3$ crystal. Further, the MR and CMR as a function of the magnetic field for all measurement configurations are shown in Figure S2 (Supporting Information), which also supports the weak magnetotransport anisotropy of the Ti:Cr$_2$Se$_3$ crystal.

Figure 3a,b show the magnetic hysteresis loops of the Ti:Cr$_2$Se$_3$ crystal with the direction of the magnetic field parallel and perpendicular to the *ab* plane of the crystal, respectively. The curves are linear above $T_N$ and "S"-like shape below $T_N$, which arises from the WFM components in low fields (< 2.5 T) and the antiferromagnetic background. The results also confirm that the in-plane magnetization is slightly stronger than the out-of-plane one at all temperatures, which is sharp contrast to that of the stoichiometric Cr$_2$Se$_3$ single crystals whose out-of-plane magnetization is ~15 times larger than the in-plane one [21]. It is thus concluded that the Ti doping is highly effective to modify the magnetic anisotropy of the Cr$_2$Se$_3$ systems. One can obtain the magnetization contributed by WFM by subtracting the antiferromagnetic background, as shown in Figure 3d where *B*//*ab*. One important characteristic of the CMR effect is the scaling of low-field MR with the square of the reduced magnetization [i.e., MR = $C(M/M_{sat})^2$], where $C$ is a constant, $M_{sat}$ is the saturation magnetization. As shown in Figure 3e, MR for the *B*//*ab* configuration in the 2–50 K temperature region are scaled by the square of the reduced magnetization. Compared with the magnetization of the stoichiometric Cr$_2$Se$_3$ crystal, the out-of-plane magnetization of the Ti:Cr$_2$Se$_3$ crystals is much weaker, however, its out-of-plane CMR (>16000%) is approximately 1143 times stronger than that of the Cr$_2$Se$_3$ single crystals (~14%). The MR and WFM component related magnetic hysteresis loops with the direction of the magnetic field perpendicular to the *ab* plane are shown in Figure 3d,f. These MR and magnetic behaviors are similar to those for the *B*//*ab* configuration (Figure 3c,d). Another point is that the *B*-perpendicular magnetic hysteresis loops at low temperatures ($T\leq20$ K) show slightly hysteresis effect, which induces the butterfly-shape MR versus *B* curves, as shown in Figure S3 (Supporting Information). Note that the butterfly-shape MR versus *B* curves were also observed in the stoichiometric Cr$_2$Se$_3$ single crystals which are absent of CMR effect [21]. Therefore, all these results demonstrate that the existing WFM should not the key reason to the CMR effect in the Ti-doped $Cr_2Se_3$ crystals.

In order to figure out the origin of the CMR effect, some classic CMR materials should be recalled. For $R_{1-x}A_x MnO_3$ manganites, CMR usually occurs near the paramagnetic-to-ferromagnetic phase transition temperature $T_C$ (usually 0.8 – 1.2 $T_C$) above which the short-range magnetic ordering as well as the magnetic polarons forms [25,26]. The presence of magnetic polarons which are charge carriers accompanied by a localized (and magnetically polarized) distortion of the surrounding crystal lattice increases the resistance of CMR materials [27]. Small angle neutron scattering study has proved the existence of magnetic polarons and its evolution with temperature and magnetic field [26], which supports the magnetic-polaron-induced CMR effect in these manganites. Similarly, CMR in other compounds like $Eu_{1-x}Gd_xSe$ [12], $Eu_5In_2Sb_6$ [16], $Eu_{14}MnBi_{11}$ [15], and $FeSb_2$ [28] occurs near the magnetic ordering temperatures. In this work, the CMR effect increases with cooling, as shown in Figure 2d, behaving very differently with the magnetic-polaron-induced CMR effect. Recently, the chiral-anomaly-induced giant negative MR in Dirac semimetals of $Cd_3As_2$ [29] and disorder-induced giant negative MR in topological insulators of $TlBi_{0.15}Sb_{0.85}Te_2$ [30] were reported, which, however, is not applicable to the Ti:$Cr_2Se_3$ crystals due to the lack of topological band structure in them.

Another interesting low-temperature CMR effect has been reported in iodine-doped transition metal dichalcogenide $TiTe_{2-x}I_x$ system whose MR reaches -85% (or CMR ~ -567%) at $B$=5 T and $T$=10 K for $x$=1. $TiTe_{2-x}I_x$ are also layered compounds which show antiferromagnet ordering at ~90 K and whose electronic transport properties depend on the iodine doping level, quite similar to those of the layered $Cr_{2+x}Se_3$ system. For $TiTe_{2-x}I_x$ crystals, CMR is due to the frustrated antiferromagnetic coupling of $Ti^{3+}$ moments as well as the band gap opening induced spin-dependent scatterings [31]. It is noted that the neutron scattering results show that the spin frustration between antiferromagnetically coupled $Cr^{3+}$ exists at low temperatures for the $Cr_2Se_3$ [19], which may become stronger upon Ti doping. The breaking of frustration by applied magnetic field would significantly reduce the resistance of Ti:$Cr_2Se_3$ crystals, resulting in colossal negative MR. As of now, the CMR effect are well known in $R_{1-x}A_x MnO_3$ manganites and some rare earth related compounds. Here, we plot the temperature and related maximum CMR values of manganites in the green area of Figure 4. While the rare earth related materials demonstrate the maximum CMR near the magnetic ordering temperatures (usually at low or extremely low temperatures), which are plotted in the pink area of Figure 4. Beyond those two materials families, some other novel materials/structures are also plotted using star symbols and shown in Table 1 (Supporting Information). The Ti:$Cr_2Se_3$ shows spin-frustration-induced CMR with a sizeable value in this diagram and may have both fundamental and applicational importance among CMR materials and inspire further investigation of this and related layered materials systems.

## 4. CONCLUSIONS

We have grown high quality layer structured $Ti_{0.023}Cr_{2.002}Se_{3.000}$ single-crystal antiferromagnets using the chemical vapor transport method and investigated the magnetic and magnetotransport properties of the crystals. A light Ti (1.5 at%) doping dramatically changes the electronic conductivity and magnetic anisotropy of the $Cr_2Se_3$ crystal from a bad metal with high magnetic anisotropy (Ref. 21) to a narrow band gap insulator with negligible magnetic anisotropy of Ti:$Cr_2Se_3$ crystal. Upon the application of a magnetic field of 14 T to the Ti:$Cr_2Se_3$ crystal with the direction of the magnetic field either parallel or perpendicular to the *ab* plane, a colossal negative magnetoresistance of ~ 16000% is achieved at 2 K, which is 1143 time larger than that of the stoichiometric $Cr_2Se_3$ single crystals without Ti doping (Ref.

21). Such a huge negative CMR is attributed to the Ti-doping enhanced frustrated spin configuration. Magnetotransport measurements demonstrate that the low-field MR at low temperatures scales with the square of the reduced magnetization. Our results demonstrate that the Ti-doping is a high effective approach to modify the electronic and magnetic properties of $Cr_2Se_3$. The Ti-doped layer structured Ti:$Cr_2Se_3$ with colossal negative CMR functionality may be used as building blocks in spintronic devices.

## ■ ASSOCIATED CONTENT

**Supporting Information**

The Supporting Information is available free of charge on the ACS Publications website at DOI: xx.xxxx/acsami.xxxxxxx. Temperature dependence of the resistance and magnetoresistance for different measurement configurations, MR and CMR versus $B$ curves for different measurement configurations, MR versus $B$ curve and magnetic hysteresis loop for $B \perp ab$ at $T$=2K.

## ■ AUTHOR INFORMATION

**Corresponding Authors**
*E-mail: zrk@ustc.edu (R. K. Zheng)
*E-mail: weiyao.zhao@monash.edu (W. Zhao)

**Author Contributions**

The manuscript was written through contributions of all authors. All authors have given approval to the final version of the manuscript.

**Notes**

The authors declare no competing financial interest.

## ■ ACKNOWLEDGMENTS

This work was supported by the National Natural Science Foundation of China (Grant Nos. 11974155). WZ and XW acknowledge the support from ARC Centre of Excellence in Future Low-Energy Electronic Technologies (CE170100039). R.K. Zheng and S.J. Zhang would like to thank Dr. Xiaoyuan Zhou and Kai Zhou from the Analytical and Testing Center at Chongqing University for their technique assistance.

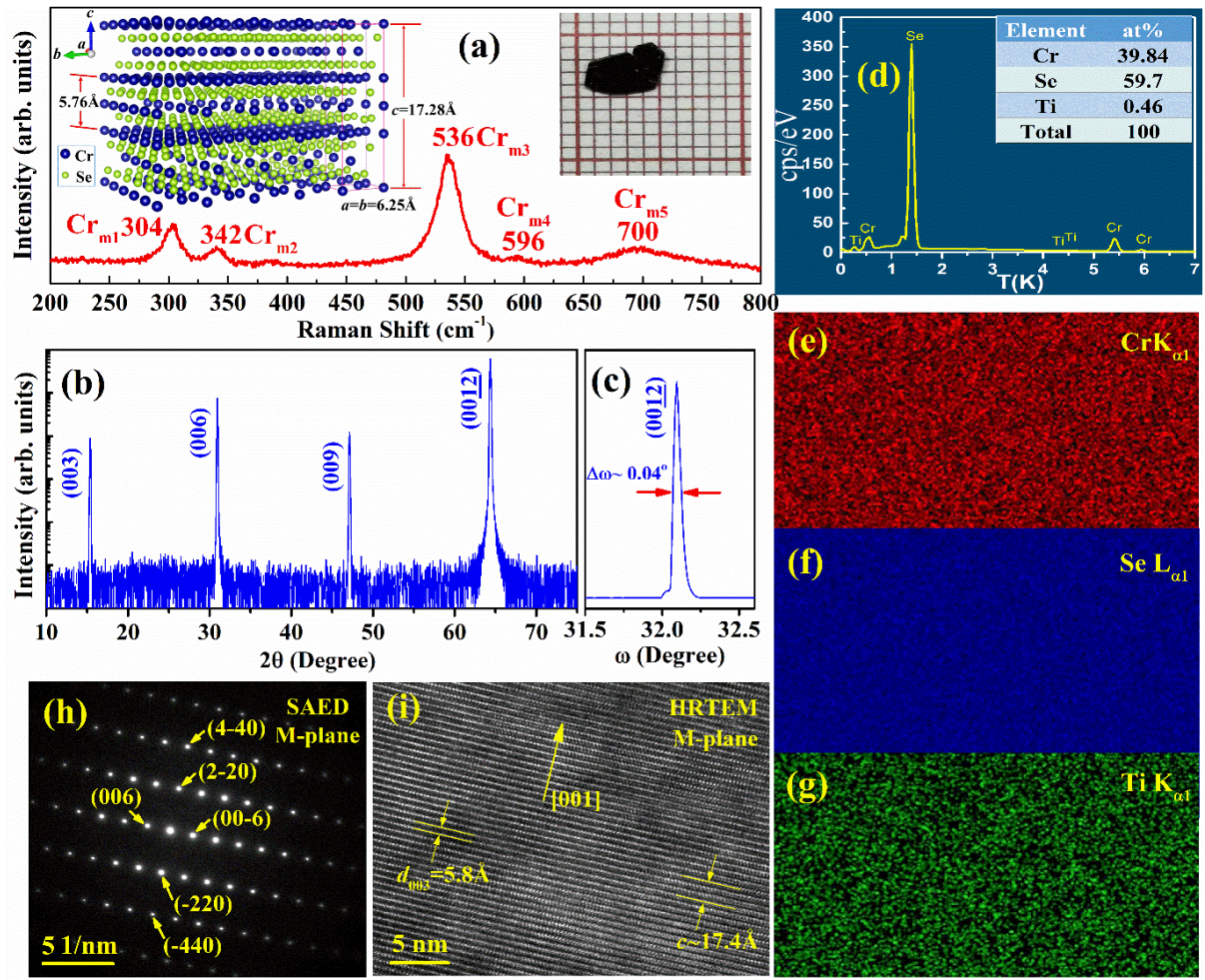

**Figure 1.** Structural and compositional characterization of a Ti:Cr$_2$Se$_3$ single crystal. (a) Raman spectroscopy. Inset: schematic crystal structure of the Cr$_2$Se$_3$ and the optical image of a Ti:Cr$_2$Se$_3$ single crystal. (b) XRD $\theta$-$2\theta$ scan pattern. (c) XRD rocking curve taken on the (00$\underline{1}$2) diffraction peak. (d) EDS results of the Ti:Cr$_2$Se$_3$ crystal, which is performed on the *ab* plane of the crystal's surface. (e-g) The area element mapping graph of the Cr, Se and Ti, respectively. (h,i) SAED pattern and HRTEM image of the Ti:Cr$_2$Se$_3$ crystal.

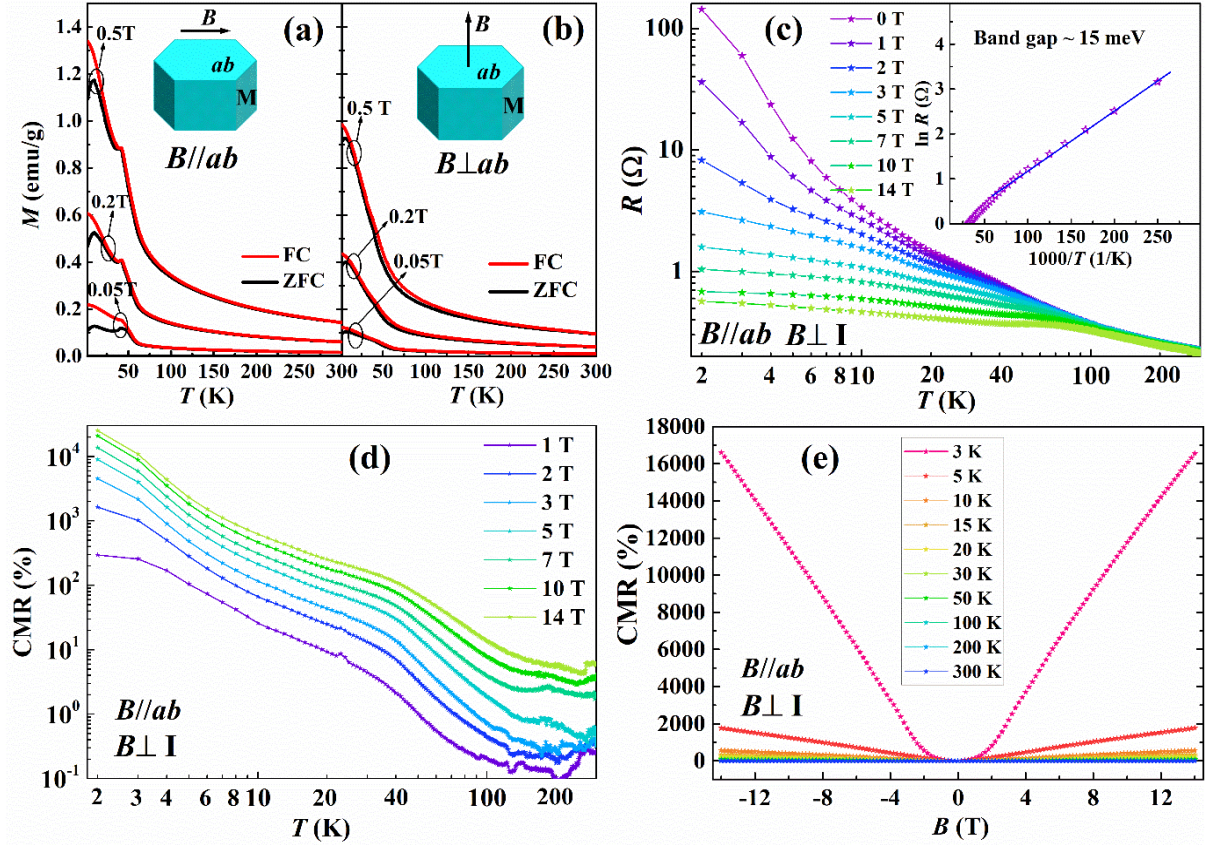

**Figure 2.** Magnetic and electronic transport properties of a Ti:Cr$_2$Se$_3$ single crystal. (a,b) Temperature dependence of ZFC and FC magnetization under different magnetic fields whose directions are parallel and perpendicular to the *ab* plane of the crystal, respectively. (c) Temperature dependence of the resistance under different magnetic fields (*B//ab*). Inset: the logarithm zero-field resistance *vs.* inverse temperature plot which is fitted linearly to demonstrate the band gap of the Ti:Cr$_2$Se$_3$ crystal. (d) Temperature dependence of the magnetoresistance under different magnetic fields (*B//ab*). (e) Magnetoresistance *vs.* magnetic field at various temperatures.

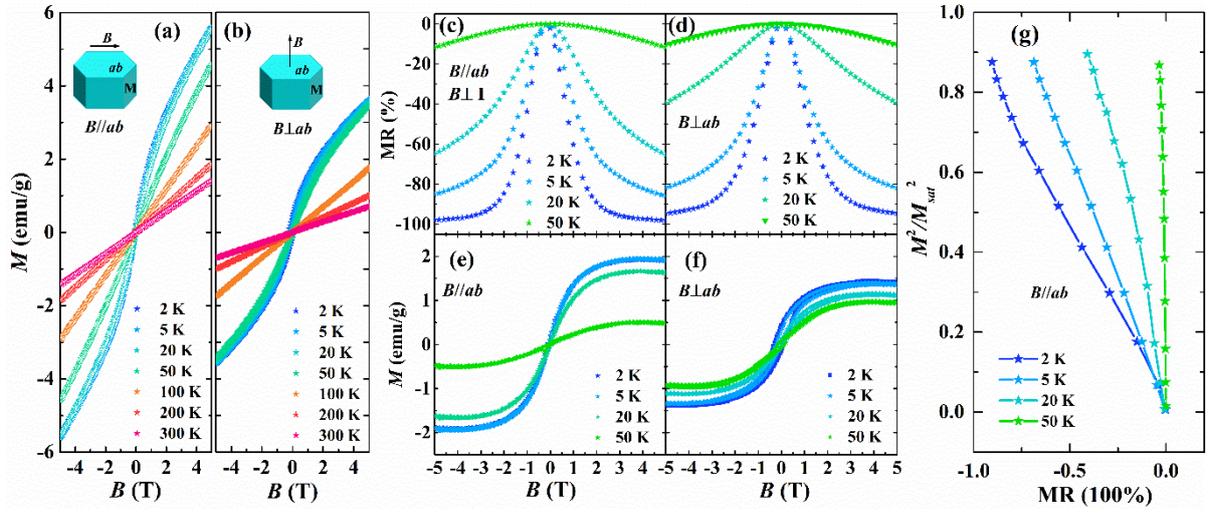

**Figure 3.** The coupling between weak ferromagnetism components and CMR effect of the Ti:Cr$_2$Se$_3$ crystal. (a, b) The magnetic hysteresis loops at various temperatures with the direction of the magnetic field $B//ab$ and $B \perp ab$, respectively. (c, d) MR versus $B$ curves at 2, 5, 20 and 50 K for $B//ab$ and $B \perp ab$, respectively. (e, f) Background-subtracted magnetization versus $B$ curves at 2, 5, 20 and 50 K for $B//ab$ and $B \perp ab$, respectively. (e) MR $vs.$ $M^2/M_{sat}^2$ plot at 2, 5, 20, and 50 K.

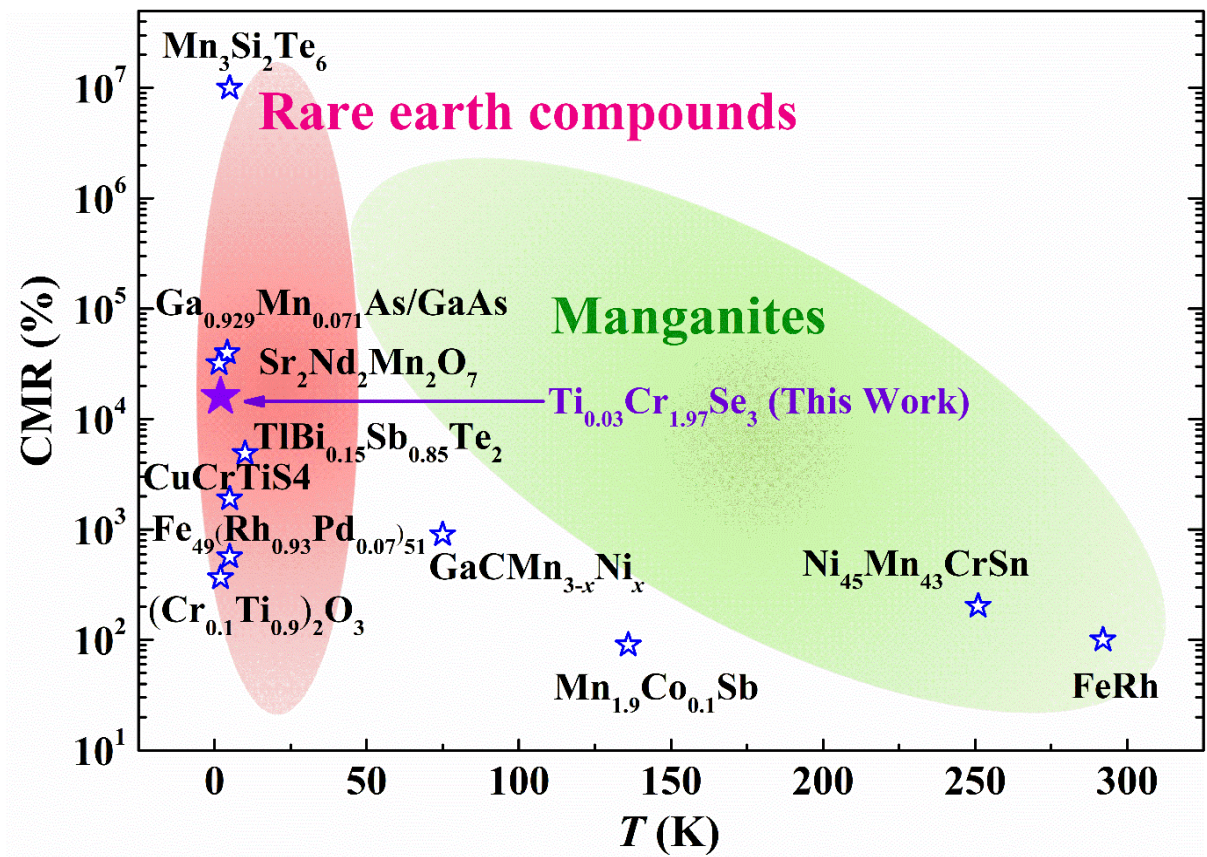

**Figure 4.** A summary of CMR values of various compounds including perovskite manganites (green area) and rare earth compounds (pink area). The detailed CMR values are shown in Table 1 of the Supporting Information.